# MASS SEGREGATION IN DARK MATTER MODELS

A. Campos[1], G. Yepes[2], A. Klypin[3], G. Murante[4], A. Provenzale[4]

and

S. Borgani[5]

[1] *Department of Physics, University of Durham, South Road, Durham DH1 3LE UK and Instituto de Astrofísica de Andalucía, Granada, SP*

[2] *Departamento de Física Teórica C-XI, Universidad Autónoma de Madrid, Cantoblanco 28049, Madrid, SP*

[3] *Astronomy Dept., New Mexico State University, Box 30001, Dept. 4500, Las Cruces, NM 88003-0001; and Astro-Space Center, Lebedev Physical Institute, Moscow*

[4] *Istituto di Cosmogeofisica del CNR, Corso Fiume 4, I-10133 Torino, IT*

[5] *INFN - Sezione di Perugia, c/o Dip. di Fisica dell'Universita', Via A. Pascoli, I-06100 Perugia, and SISSA, Trieste, IT*

## ABSTRACT

We use the moments of counts of neighbors as given by the Generalized Correlation Integrals, to study the clustering properties of Dark Matter Halos (DH) in Cold Dark Matter (CDM) and Cold+Hot Dark Matter (CHDM) models. We compare the results with those found in the CfA and SSRS galaxy catalogs. We show that if we apply the analysis in redshift space, both models reproduce equally well the observed clustering of galaxies.

Mass segregation is also found in the models: more massive DHs are more clustered compared with less massive ones. For example, at $3h^{-1}$ Mpc in real space, the number of neighbors in excess of random increases $\sim 3$ times when the mass of halos increases by $\sim 100$. In redshift space, this mass segregation is reduced by a factor $\sim 2-3$ due to the peculiar velocities. Observational catalogs give an indication of luminosity and size segregation, which is consistent with the predictions of the models. Because the mass segregation is smaller in redshift space, it is suggestive that the real luminosity or size segregation of galaxies could be significantly larger than what it is found in redshift catalogs.

*Subject headings:* cosmology: theory - dark matter - large-scale structure of universe -galaxies: clustering

# 1. INTRODUCTION

During the last years, a major effort has been devoted to the analysis of the galaxy distribution in the Universe as a direct test of the different cosmological models. On a purely statistical ground, there is still controversy about to what extent present-day catalogs might be considered as *fair* samples of the Universe. In fact, there is a general agreement that larger and deeper catalogs are needed. Besides, many authors have pointed out that a better understanding of the galaxy formation process is strongly needed. In fact, using observational samples we can only analyze the distribution of *visible matter in the form of galaxies*, and from that we can try to extract information about the actual distribution of matter (visible and non-visible, baryonic and non-baryonic) in the Universe. This can be a very compelling task as long as we do not have an enough insight into the physics of galaxy formation and evolution.

In this situation, a crucial problem is related to the possible differences that could well exist between the distributions of galaxies and dark matter and even between different types of galaxies. Kaiser (1984) introduced the concept of *bias* by showing that the correlation function of peaks is larger than that of the underlying matter distribution by some factor $b$, called the linear bias parameter.

The actual biases in the galaxy distribution could be much more complicated than in this simple picture. An interesting problem related to these biases is the segregation of galaxies as a function of their mass, morphology, luminosity, etc. That is, galaxies could not only be bad tracers of the mass distribution in the Universe but, even worse, *bright* galaxies could not even be good tracers of the distribution of *all type of* galaxies in the Universe. In fact, it has been suggested that any suitable model of biased galaxy formation would produce segregation of galaxies (Dekel & Rees 1987). For example, in the standard cold dark matter (CDM) model it is predicted that objects with high circular velocities should populate the highest density regions (White et al. 1987).

Luminosity segregation of relatively bright galaxies is not large. But it has been found by several authors using different statistical techniques (i.e. Davis et al 1988; Haynes et al 1988; Hamilton 1988; Tully 1988; White, Tully & Davis 1988; Börner, Deng & Xia 1989; Börner & Mo 1989, 1990; Salzer, Hanson & Gavazzi 1990; Santiago & da Costa 1990; Maurogordato & Lachieze-Rey 1991). The presence of luminosity segregation has been reconfirmed by the recent analysis of the CfA 2 by Geller et al. (1994), who show that segregation does exist at a $2\sigma$ confidence level. On the other hand, morphological segregation was found between elliptical and spiral galaxies, with the former appearing mainly in rich groups and clusters whereas the latter populate the general field (Davis & Geller 1976; Dressler 1980; Giovannelli, Haynes & Chincarini 1986; Santiago & Strauss 1992; Dominguez-Tenreiro et al. 1994). Iovino et al. (1993) have shown that the two types of segregation are separated effects with morphology being more strongly dependent on the clustering properties.

In the present work, we study the mass segregation in CDM and CHDM models by means of the moments of counts of neighbors, (Scaling analysis). The scaling analysis method was first applied to the large scale galaxy distribution by Jones et al. (1987; see also Martínez et al. 1990). A detailed scaling analysis of the distribution of galaxies in the Center for Astrophysics Redshift Survey (CfA) (Huchra et al 1983) and in the Southern Sky Redshift Survey (SSRS) (da Costa et al. 1991) has been recently reported in Domínguez-Tenreiro et al. (1994; hereafter DGM) and in Campos et al. (1994; hereafter CDY). In these two works, the so-called *Generalized Correlation Functions* (Hentschel & Procaccia 1983) as well as the *Density Reconstruction Method* (Grassberger, Badii & Politi 1988) were used as a full set of statistical descriptors. There is a main advantage of using these descriptors. They are capable of probing regions of different density because they contain high order correlations in their definition. See Borgani (1995) for a recent review about scaling analysis methods for large scale structure studies.

Murante et al (1994; hereafter MPBCY), have made an exhaustive analysis of the clustering properties of Dark Halos (DHs) in high resolution N-body simulations of CDM and CHDM models by means of the multifractal methodology. One aim of the present work is to make a direct comparison between the numerical models and the CfA and SSRS catalogs.

We show that, within the errors, both CDM and CHDM models reproduce the observed properties of the optical galaxies. The most important result we report so far is the possible existence of mass segregation in the dark halo (DH) population. The amount of segregation is qualitatively and even quantitatively similar to the luminosity, size and morphological segregation observed in the catalogs. We also show that, in redshift space, this segregation can be up to 3 times smaller than in real space due to the effect of the peculiar velocities.

This paper is organized as follows: In §2 we give a brief description of the statistical method used in the present work. In §3 we present the galaxy and DH data sets used in the statistical analysis. The main results of this work are reported in §4. Section 5 is devoted to the discussion and conclusions.

## 2. STATISTICAL DESCRIPTORS

In this section we give a brief description of the statistical functions used in the following analysis. More detailed information on the multifractal methodology can be found elsewhere (Paladin & Vulpiani 1987; Martínez et al 1990; Borgani et al 1993; Borgani 1995).

In a point set one defines a probability measure for each point of the set as:

$$p_i(r) = \frac{n_i(r)}{N}, \qquad (1)$$

where $n_i(r)$ is the number of points inside a sphere of radius $r$ centered at the point $i$ of the distribution and $N$ is the total number of points of the set.

The correlation integrals, $Z(q,r)$ are defined as

$$Z(q,r) \equiv \frac{1}{N} \sum_{i=1}^{N} p_i(r)^{q-1} \quad , \quad (q \neq 1) \quad (2)$$

$$Z(q,r) \equiv \frac{1}{N} \sum_{i=1}^{N} p_i(r) \log p_i(r) \quad , \quad (q = 1). \quad (3)$$

Depending on the value of $q$, different regions of the set are weighted in different ways. For high positive $q$, the sum in Eq (2) is dominated by points in high density regions, while for high negative $q$, the more rarefied regions of the distribution are statistically dominant. This is one of the main advantages of using this technique, as one gets an estimate of the statistical properties of a point set for regions of different local density.

For a better comparison with the standard correlation function analysis, we define the quantity

$$J(q,r) = \frac{Z(q,r)}{Z_{\mathrm{rand}}(q,r)} \qquad (4)$$

$$Z_{\mathrm{rand}}(q,r) = \left(\frac{4\pi r^3}{3V}\right)^{q-1} \qquad (5)$$

with $V$ being the total volume of the sample. The $J(q,r)$ functions describe the *excess* of clustering, as measured by $Z(q,r)$, with respect to a random distribution. Therefore for $q = 2$, $J(2,r) \propto (1 + J_3(r)/r^3)$, where $J_3(r)$ is the well known integral of the two-point correlation function. For other values of $q$ it is not possible to give a simple general relationship with the higher-order correlation functions, although all the N-point correlation functions are implicitly included in the definition of $J(q,r)$ (see eg. Balian & Schaeffer 1989, Borgani 1993).

For negative values of $q$, the sum in Equation (2) is dominated by regions with a small number of points. Thus it has poor statistics and is more noisy. In this case, an alternative method has been proposed which is better suited to study low density regions. It is commonly called the *density reconstruction method* (Grassberger, Badii & Politi 1988) and is based on the computation of the moments

$$W(\tau,p) \equiv \frac{1}{N} \sum_{i=1}^{N} r_i(p)^{-\tau} \qquad (6)$$

where $r_i(p)$ is the radius of the smallest sphere centered at the point $i$ which contains $n = pN$ points.

The above formulation is well suited for the study of the observed galaxy distribution. To make a comparison with numerical models, one has to find *galaxies* in the simulated dark matter distribution. As discussed in MPBCY, this may be a problem because of the overmerging of DH's due to the lack of resolution (see also below). A way to overcome this problem, and the one followed here, is to weight the DHs by their mass. In this way, we analyze the statistical properties of the mass distribution of the DHs instead of those of the DH number density. A straightforward way of doing this is to generalize the probability measure, $p_i(r)$, assigned to each point in the set as

$$p_i(r) = \frac{\sum_{j=1}^{N} w_j \, \theta(|\mathbf{x}_i - \mathbf{x}_j| - r)}{\sum_{j=1}^{N} w_j} \qquad (7)$$

where $w_j$ is the mass associated with the point $i$ and $\theta$ is the Heaviside step function. With this definition of the probability measure we can estimate $Z$ and $W$ for the mass distribution (i.e. *mass weighted*) and compare with the results of $Z$ and $W$ for the point distribution (i.e. *unweighted*), corresponding to assume $w_i = 1$ for all $i$.

According to equations 2 and 6, the more clustered the point (or mass) distribution is, the shallower the $Z(q,r)$ and the $W(\tau,p)$ curves will be. A direct comparison of the $Z$ and $W$ functions of one data set with $Z$ and $W$ of another data set will thus provide valuable information on the relative difference in the clustering properties as a function of scale and for different parts of the distributions.

## 3. GALAXY AND DARK HALO DATA SETS

To compare the numerical models with the observational data, we use the CfA and SSRS optical redshift catalogs. The CfA contains redshifts of all the galaxies with $\delta \geq 0°$ and Galactic latitude $b^{II} \geq 40°$, with a Zwicky blue magnitude brighter than 14.5. The SSRS compiles the redshifts of all galaxies located in an area defined by $\alpha \in [5^h - 19^h]$, $\delta \leq -17°$ and $b \leq -30°$, with a major optical apparent diameter larger than 1.26 arcmin. These two catalogs have been analyzed by DGM and CDY using the scaling formalism. In these two works, 5 volume-limited samples were studied for each catalog. The samples had a minimum limit in radial velocity (namely, 1700 km s$^{-1}$ and 2000 km s$^{-1}$ for the CfA and SSRS respectively) to avoid the rich galaxy clusters of Virgo and Fornax, whose presence in the volume-limited samples could dominate the statistical results. In what follows we will only show results for samples with a depth of 50 h$^{-1}$ Mpc as representatives of the galaxy distribution in the catalogs, although several other sub-samples, corresponding to different depth, have been considered as well. Throughout this paper the sub-samples will be called galaxy data sets. For further details on the number density of objects, sample volumes and other characteristics of the galaxy data sets we refer to DGM and CDY.

As far as the numerical models are concerned, we have used the same set of high-resolution numerical simulations of CDM and CHDM discussed by Nolthenius, Klypin and Primack (1993), which have been extensively analyzed in MP-BCY.

The CHDM model analized here correspond to $\Omega_{cold} = 0.6$, $\Omega_{hot} = 0.3$ and $\Omega_{baryons} = 0.1$, and the initial spectrum (Klypin et al. 1993) is normalized to the COBE quadrupole of 17 $\mu K$, corresponding to $\sigma_8 = 1/1.5$ for the r.m.s. fluctuations within a top-hat sphere of radius $8h^{-1}$Mpc. The same $\sigma_8$ value is also used for normalizing the CDM power spectrum. The simulations were carried out using a particle-mesh code on a $512^3$ mesh in a 100 Mpc box ($H_0 = 100h$ km s$^{-1}$ Mpc$^{-1}$, $h = 0.5$).

To compare observations with numerical models it is necessary to make an assumption about how galaxies are formed and how can they be identified in the models. Several prescriptions have been used by different authors, with varying degrees of complexity. In the present analysis a simple overdensity algorithm is used. All local maxima on the simulation mesh are found. Only those with overdensity larger than 100 are retained for the analysis. Then we assign to each halo the mass inside the cell and the total mass of its 26 neighbors. This gives the mass limit for halos of $\sim 3 \times 10^{11} M_\odot$. Note that the efective radius of the halos is about 180 kpc. Because this is $10-20$ times bigger than the radius of galaxies, one expect that the mass-to-light ratio should be $\sim 200$, which gives an effective magnitude limit for DHs of about -17.5. As a result of this, we have much more halos in the simulations as compared with the observations, where the typical magnitude limit was $\sim -19$. In order to make a direct comparison between the simulated and real catalogs, it is necessary that both have the same mean density. To do that, we have randomly selected from the simulation catalogs the same number of points than in the real catalogs. This procedure provides us also with a direct estimation of the statistical errors.

We also recall that there is a well-known problem of "overmerging" due to the lack of resolution in numerical N-body simulations. In high density regions, where a large number of individual DHs is expected, the actual number of halos is relatively small because some of them were artificially merged and produced too big DHs. For some purposes (like the two-point correlation function) one can overcome this problem by weighting the DHs by their masses (Klypin, Nolthenius & Primack 1993; Bonometto et al 1995; see also MPBCY). This overmerging results in mass segregation: largest DHs are found in centers of groups and clusters. The problem, nevertheless, is not as severe as it used to be because of the improved force and mass resolution of the simulations. At the same time we note that in the Universe the most massive galaxies are also often found in groups or clusters. As we show here, the difference between observationally allowed segregation and what is observed in the simulations of the CHDM model is not large.

To make a reliable comparison between the results obtained from the galaxy catalogs and those provided by the numerical models, it is then necessary to obtain simulation subsamples with the same characteristics (volume, shape, density) as the real catalogs. To this end, we have extracted from the simulations several DH data sets which are identical in volume, shape and number of objects to the galaxy data sets extracted from the catalogs. This has been made with the aim of ensuring that the discrepancies that could eventually exist between the distributions of galaxies and that of the DHs were not produced by any statistical artifact due to differences in the number of particles, boundary corrections etc. In fact, we have applied the same corrections for boundary effects to the $Z$ and $W$ partition functions in both cases. In this work we have used the same boundary correction which has been employed by DGM and CDY in the analysis of the galaxy catalogs and by Yepes et al (1992) in the analysis of low bias CDM simulations. We refer to this last paper for further details.

To extract the DH data sets from the simulations, we have used the following procedure. We have replicated the periodic cubic box of the simulation 27 times (3×3×3), placed the observer at the center of this larger box and then calculated the redshifts that such an observer would measure (i.e. cosmological plus peculiar velocities). Then, we have extracted different cones which are identical in shape to those surveyed by the two catalogs. For each cone, DH data sets which were identical in shape and number density to the galaxy data sets analyzed in DGM and CDY were considered for the comparison.

This procedure was repeated several times, placing the cones with different orientations in order to cover most of the simulation volume. It should be pointed out that, although we replicated the cubic box, there are no duplicated structures in the DH data sets that could mask the results, since we never took twice the same region of space. Finally, we calculated the average values for the $Z$ and $W$ functions as well as the standard deviations. In this work, we show results by averaging over 14 different cones and the error bars represent the $2\sigma$ deviation from the average.

As it is shown in Tables 1 of DGM and CDY, the number of galaxies in the observational data sets is relatively low ($\sim 4h^3 \times 10^{-3}$ gal/Mpc$^3$) as these are volume-limited samples. However, as it is also shown in these papers, the statistical errors (estimated with the standard bootstrapping technique) are very small ($\sim 14\%$ at $1h^{-1}$ Mpc). Larger errors could however be expected due to inhomogeneities in the galaxy distribution. Because the volumes of the data sets are relatively small, the results could strongly depend on the presence or absence of structures. Nevertherless, it has been shown (DGM and CDY) that the results for the different galaxy data sets are fairly consistent. As for the DH data sets, the error bars shown in this paper are a measure of the inhomogeneities in the distribution of DHs in the simulations and they are always larger than the expected statistical errors as evaluated by the bootstrap technique. For instance, an estima-tion of the cosmic variance from our simulations gives a relative error $\Delta Z(2,r)/Z(2,r) \sim 40\%$ at $r = 1 h^{-1}$ Mpc, which is more than 4 times bigger than the bootstrap error in the catalogs.

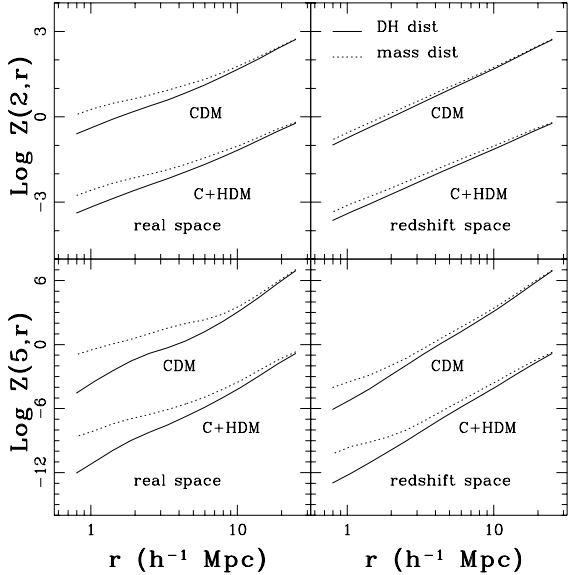

Fig. 1.— Unweighted (solid lines) and mass-weighted (dashed lines) $Z(q,r)$ computed for CDM and CHDM Dark Halo distributions in redshift and real space. The two $q$-moments shown here correspond to moderate ($q = 2$) and high ($q = 5$) density areas in the distributions. The $Z$ values have been rescaled for a better comparison between the two models.

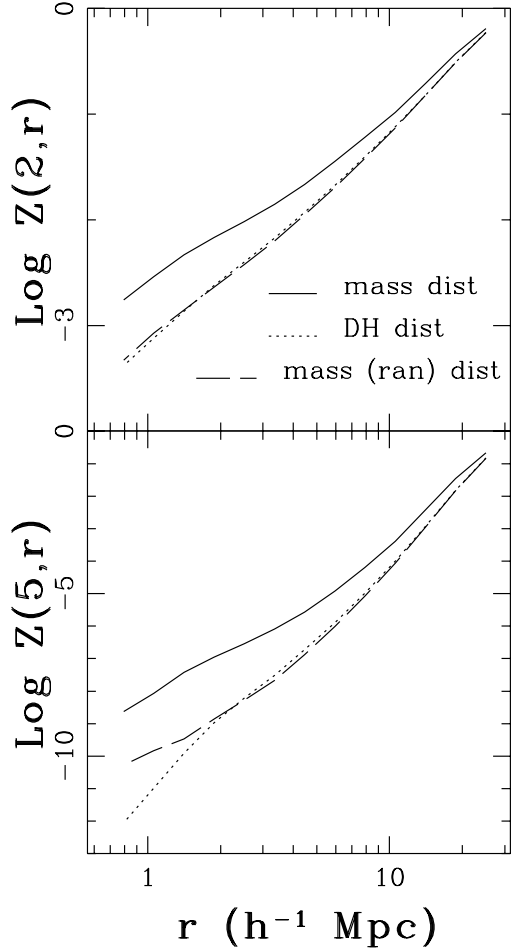

Fig. 2.— $Z(2,r)$ and $Z(5,r)$ for the CHDM Dark Halo distribution in the periodic cubic box of the simulation. The functions have been computed in real space with three different methods: mass-weighted (solid line), number weighted (dotted line) and mass-weighted after a random distribution of DHs masses (dashed-lines).

## 4. RESULTS

The first problem explored here is the smoothing in the DH distribution induced by the lack of resolution in the simulations. In Figure 1 we show the $Z$ functions for $q = 2$ and $q = 5$ as computed from the distribution of DHs in real and redshift space, both mass-weighted and number weighted. A clear difference between the mass-weighted and number weighted $Z$ functions is present. This difference is seen both in real and redshift space. The DH distribution is less clustered when the number-density distribution is considered.

The DH mass-density and DH number-density distributions should have the same clustering properties and therefore there should be no differences in the $Z$ functions. As it can be seen in Figure 2, this is the case, indicating that the differences between mass and number distribution of DHs is a real effect. We also note that the difference between the DH mass-density and DH number-density distributions is smaller by a factor of $2-3$ in redshift space than in real space.

To check that our algorithm does not produce spurious effects, we randomly re-distributed the mass of the DHs. The DH mass-density and DH number-density distributions should have the same clustering properties and therefore there should be no differences in the $Z$ functions. As can be seen in Figure 2, this is the case, indicating that the differences between mass and number distribution of DHs is a real effect. We also note that the difference between the DH mass-density and DH number-density distributions is smaller by a factor of $2-3$ in redshift space than in real space.

In Figures 3 and 4 we plot the number weighted $Z$ ($q = 2$ and $q = 5$) and $W$ ($\tau = -1$ and $\tau = -5$) functions for the DH number-density in CDM and CHDM models, obtained from the DH data sets equivalent (in volume, shape and number of objects) to the CfA and SSRS data sets with a depth of 50 $h^{-1}$ Mpc. The results for the galaxy data sets as given by CDY and DGM are shown as well. It can be seen that, despite the overmerging problem, the fit of both models (CDM and CHDM) to the observational samples is reasonably good, and it does not allow for a clear selection of one of the two models.

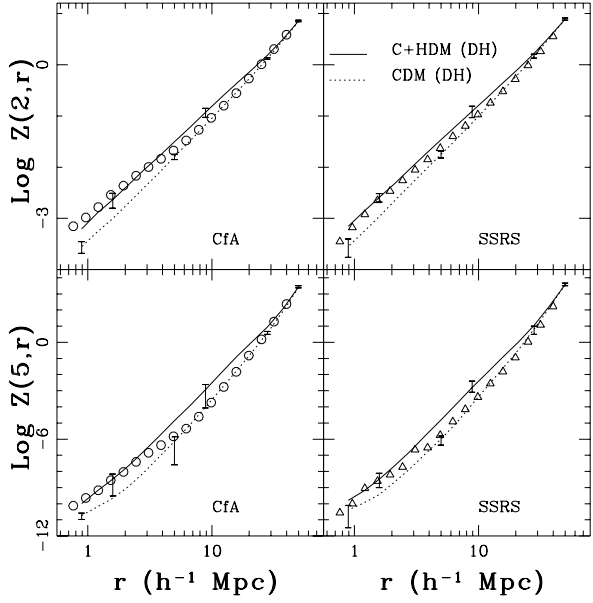

Fig. 3.— $Z(2,r)$ and $Z(5,r)$ functions (unweighted) for the CDM (dotted lines) and CHDM (solid lines) Dark Halo distribution in redshift space. The curves correspond to the average of the $Z$ functions computed in a set of 14 different volumes, equivalent to the CfA and SSRS galaxy catalogs respectively. Error bars correspond to $2\sigma$ deviation from the average. For comparison we also plot results for the real catalogs taken from DGM and CDY.

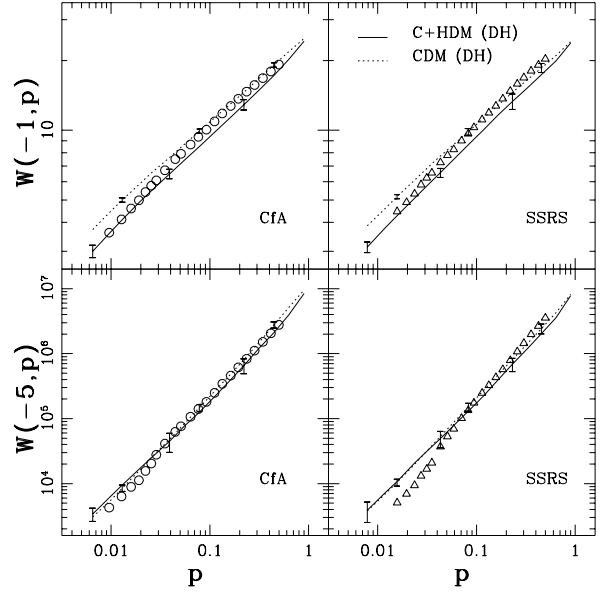

Fig. 4.— Same as Fig 3, but for $W(-1,p)$ and $W(-5,p)$ as a function of probability.

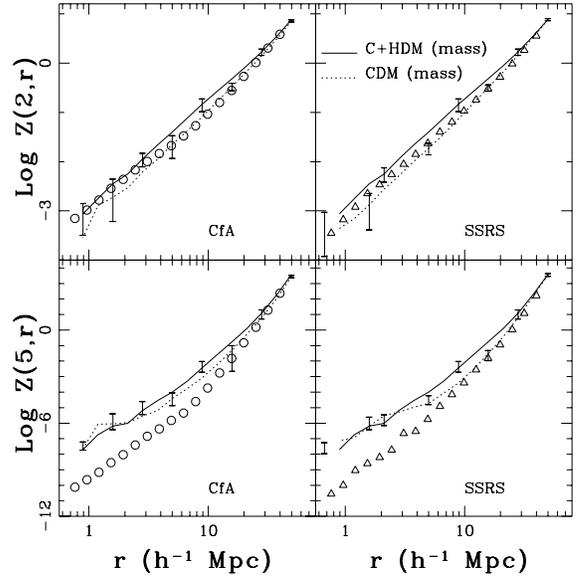

Fig. 5.— Same as in Fig 3, but here the $Z$-functions are computed weighting by mass the DH's.

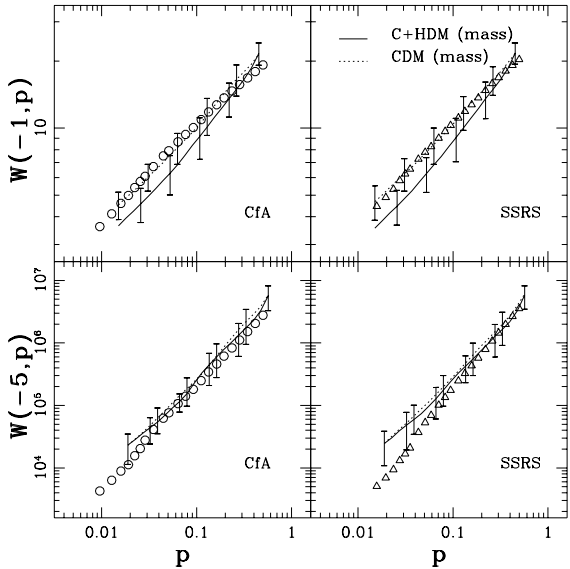

Fig. 6.— Same as Fig 4, but with $W$–functions computed by mass-weighting the DHs.

In Figures 5 and 6 we show the results of the same analysis, but for the DH mass-density distribution (i.e. mass-weighted $Z$). The results for the galaxy distribution and the DH distribution for large values of $q$, where the role of the high density areas is predominant, are different. The disagreement between the observational data, (for which the galaxy number-density distribution is analyzed), and the numerical models (DH mass-density distribution) could be explained as due to a real physical segregation of objects in the models as a function of their mass.

To test whether we see a similar effect in the galaxy data sets, we have computed the $Z$ functions for the CfA samples, by weighting the galaxies by their luminosities (in the B band). In Figure 7 we plot the results for the galaxy number-density and luminosity-density distributions. It can be seen that there is a difference between both distributions, although smaller than the one shown between the galaxy number-density and the DH mass-density.

## 5. DISCUSSION AND CONCLUSIONS

The comparative analysis of the CDM and CHDM simulations and of the galaxy catalogs has led to two main results. The first one is the absence of clear differences between the CDM ($b = 1.5$) and CHDM models. On $1 - 10h^{-1}$ Mpc scales, CHDM provides stronger clustering than CDM as it can be seen in Figures 3–6. These differences are not large enough to rule out any of them when comparing with the observational data. A better distinction between the two models is, however, provided by the scaling properties at larger scales, (see MPBCY).

The second results is the presence of mass segregation in the models. A direct comparison between the number-density and mass density distributions of the DHs reveal its presence. This is a common effect of all N-body simulations due to the overmerging problem previously mentioned. When a comparison between number density of DHs and number density of galaxies in the catalogs is made (Figs 3-4) we observe a very good agreement. This means that the number density distribution of DHs in the models traces reasonably well the observed galaxy distribution. When we compare the mass-density distribution (Figs 5-6) with the same galaxy distribution, we observe a statistically significant difference between the galaxy distribution and the distribution of halos located in the high density regions (as described by the mass-weighted $Z(5,r)$); i. e. when we correct from the overmerging problem (weighting by mass) we get no agreement with the observations at all. This would imply that the overmerging in the simulations is, in fact, necessary if one wants an agreement with the observational data. Therefore, it would be possible that the overmerging of DHs in the high density regions could well be a gravitational effect and not a numerical artifact as it has been considered until now. In recent numerical simulations with the inclusion of cooling effects and supernova explosions (Yepes et al 1994) the expected breaking of the massive halos into smaller ones is not observed. This would also support the idea that the mass segregation is the natural result of non linear gravitational evolution. On the other hand, the existence of mass segregation is consistent with the multifractal nature of the dark matter density field at small scales, as discussed in Valdarnini et al (1992) and in Yepes et al (1992). The above results also imply that, if galaxies of different mass have different clustering properties, the statistical analysis of the number density of galaxies may hardly be compared with that of the numerical simulations, unless a suitable, physically motivated mechanism for galaxy formation is provided.

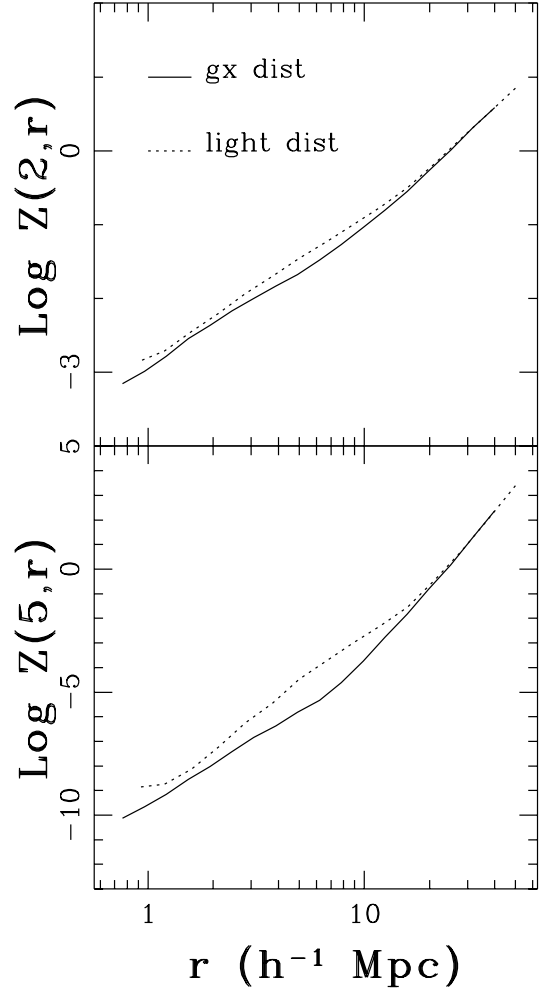

Fig. 7.— $Z(2,r)$ and $Z(5,r)$ computed for the same CfA sample than in Figs 3-6 (see text). Solid line correspond to $Z$ computed for the galaxy distribution (unweighted). Dotted lines correspond to $Z$ computed by weighting the galaxies in the sample according to their absolute luminosity.

The mass segregation found in the models could be related to the luminosity segregation observed in the CfA (see Figure 7) and/or with the size segregation observed in the SSRS catalog (CDY). In fact, the luminosity segregation as seen in Figure 7 is actually a lower limit of the real mass segregation in the galaxy distribution. If the M/L ratio had the same value for all galaxies, then the mass-density and the luminosity-density distributions would be exactly the same. It is well known, however, that the M/L ratio varies along the Hubble sequence, ranging from values around 1 in early types up to 10 or more in late types. On the other hand, there are evidences that elliptical galaxies are more clustered than spiral galaxies, up to scales of 10-15 $h^{-1}$ Mpc (Santiago & Strauss 1992; Mo et al. 1992; Domínguez-Tenreiro et al. 1994). The fact that the elliptical galaxies have stronger correlations while having larger M/L ratio than the spiral galaxies makes this luminosity segregation to be a lower limit of the real mass segregation that could well exist between galaxies. In this regard, Iovino et al. (1993) have claimed that, even if luminosity and morphology seem to be independent parameters in the galaxy distribution, the latter has a predominant role. They argue that morphology could well be coupled with mass, perhaps better than luminosity, by a mechanism that still remains to be elucidated. In Figure 8 we plot $J(2,r)$ for the number-density and luminosity-density distributions from the CfA data set. Luminosity segregation is clearly observed up to scales of 10 $h^{-1}$ Mpc, as already reported by several authors.

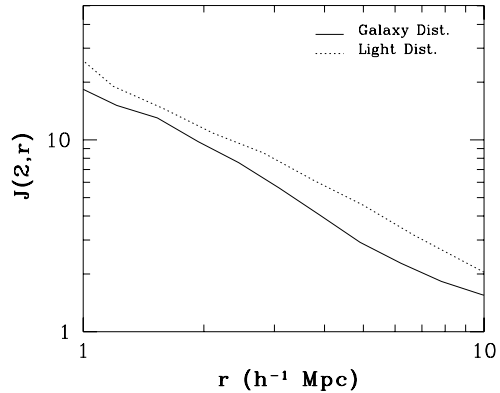

Fig. 8.— $J(2,r)$ for the same CfA sample than in Fig 7 and for the luminosity-weighted (dashed line) and unweighted (solid line) schemes. $J(2,r)$ values represent the *excess* of clustering with respect to a random distribution as measured by $Z(2,r)$ and is related to the $J_3(r)$ integral of the two-point correlation function.

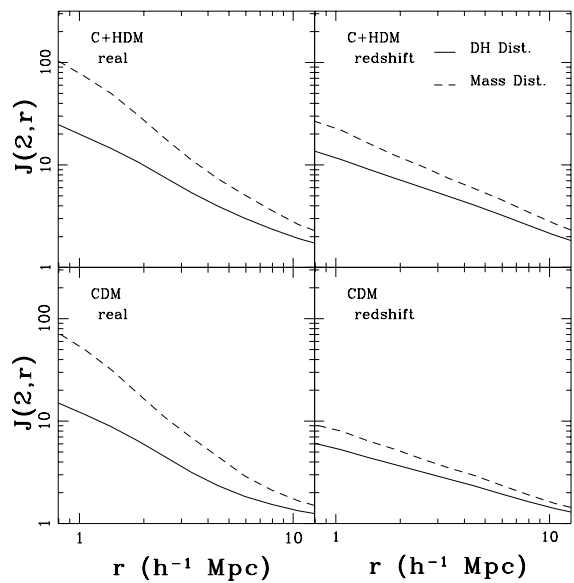

Fig. 9.— Mass weighted (dashed lines) and unweighted (solid lines) $J(2,r)$ functions for the Dark Halo distribution in the numerical models. Results in real and redshift space are shown for comparison.

It must be taken into account that galaxies are observed in redshift space, where peculiar velocities can severely distort the structures. In fact, any existing segregation would be masked by the peculiar velocities of the halos in the more dense areas. To estimate how big this effect is, we plot in Figure 9 $J(2,r)$ from the DH number-density and mass-density distributions, both in real and redshift space. It is clearly seen in this Figure that the differences are smaller in redshift than in real space. For instance, at a scale of $5h^{-1}$ Mpc the *dilution* factor is of the order of 2 in CHDM, and of the order of 3 in CDM.

In general, the results discussed in this work support the existence of biasing in the galaxy formation process, in the sense that galaxies would form in preferred places according to, in principle, their masses. Consequently, galaxies of different mass would trace the density field in a different way, making the standard linear bias picture an oversimplified approach.

GY wishes to thank CICyT (Spain) for financial support under project number PB90-0182.

---